# Synthesis and Properties of *c*-axis Oriented Epitaxial $MgB_2$ Thin Films

**S.D. Bu[a)], D.M. Kim [a)], J.H. Choi[a) b)], J. Giencke[a)], S. Patnaik[b)], L. Cooley[b)], E.E. Hellstrom[a) b)], D.C. Larbalestier[a) b)] and C.B. Eom[a) b) c)]**
[(a)] Department of Materials Science and Engineering
[(b)] Applied Superconductivity Center
University of Wisconsin-Madison, Madison, Wisconsin 53705

**J. Lettieri, D.G. Schlom.**
Department of Materials Science and Engineering, Pennsylvania State University, University Park, Pennsylvania 16802

**W. Tian, X. Q. Pan**
Department of Materials Science and Engineering, University of Michigan, Ann Arbor, Michigan 48109

We report the growth and properties of epitaxial $MgB_2$ thin films on (0001) $Al_2O_3$ substrates. The $MgB_2$ thin films were prepared by depositing boron films via RF magnetron sputtering, followed by a post-deposition anneal at 850°C in magnesium vapor. X-ray diffraction and cross-sectional TEM reveal that the epitaxial $MgB_2$ films are oriented with their *c*-axis normal to the (0001) $Al_2O_3$ substrate and a 30° rotation in the *ab*-plane with respect to the substrate. The critical temperature was found to be 35 K and the anisotropy ratio, $H_{c2}^{\parallel} / H_{c2}^{\perp}$, about 3 at 25K. The critical current densities at 4.2 K and 20 K (at 1 T perpendicular magnetic field) are $5x10^6$ A/cm$^2$ and $1x10^6$ A/cm$^2$, respectively. The controlled growth of epitaxial $MgB_2$ thin films opens a new avenue in both understanding superconductivity in $MgB_2$ and technological applications.

The discovery of superconductivity at 39 K in $MgB_2$[1] offers the possibility of a new class of high-speed superconducting electronic devices due to its favorable combination of higher critical temperature than conventional BCS superconductors and a symmetric order parameter (unlike HTS). It also stimulated a flurry of activity to explore the phenomenology and basic mechanism of superconductivity in this surprising material. $MgB_2$ possesses a number of attractive properties, including strongly coupled grain boundaries[2]. Several unusual phenomena, such as temperature-dependent electronic anisotropy[3] and multiple superconducting gap structures[4,5], appear to distinguish $MgB_2$ from a conventional BCS superconductor, and remain to be explained.

A critical step for studying both intrinsic superconducting properties and the possibility of superconducting devices based on $MgB_2$ is the controlled growth of high quality epitaxial $MgB_2$ thin film heterostructures. The growth of $MgB_2$ films by means of both *in-situ* and *ex-situ* processes has been demonstrated[6-12], including (0001) fiber-textured $MgB_2$ films[7]. Kang *et al*. reported the growth of both *c*-axis and (101)-oriented $MgB_2$ epitaxial thin films on $(1\bar{1}02)$ $Al_2O_3$ and (100) $SrTiO_3$ substrates[6]. However, the reported x-ray data do not show in-plane epitaxy, and there is no clear relationship between the $MgB_2$ film orientation and the orientation of the substrate, which must be present if epitaxial control over the film growth had been attained.

In this letter, we report the growth and properties of epitaxial $MgB_2$ thin films. The films were grown on a variety of substrates, including (111) $SrTiO_3$, (001) MgO, (111) MgO, (001) $MgAl_2O_4$, (110) $MgAl_2O_4$, (111) $MgAl_2O_4$, and (0001) $Al_2O_3$. Among these substrates, (0001) $Al_2O_3$ promotes the strongest epitaxial growth. Although $MgB_2$ has a mismatch with $Al_2O_3$ of ~46% along the $[2\bar{1}\bar{1}0]$ axis, which is unfavorable for epitaxial growth, a 30° in-plane rotation of the $[2\bar{1}\bar{1}0]$ direction of the $MgB_2$ film with respect to the substrate results in a parallel orientation of $[2\bar{1}\bar{1}0]$ $MgB_2$ and $[10\bar{1}0]$ $Al_2O_3$. This provides a lattice mismatch of ~9%.

The $MgB_2$ thin films were prepared by depositing boron via RF magnetron sputtering, followed by a post deposition anneal at 850°C in the presence of magnesium vapor. Two different epitaxial $MgB_2$ thin films on (0001) $Al_2O_3$ are discussed in this letter. The base pressure before boron deposition was $3x10^{-6}$ Torr. Deposition was carried out at 5 mTorr argon at 500°C using a pure boron target. The thickness of the boron films was 230 nm. The films were annealed in an evacuated quartz tube using a tantalum envelope. The quartz tube was filled with 7-10 Torr of Argon gas after evacuation to reduce the Mg loss. Film 1 was annealed for 30 minutes, while Film 2 was annealed for 5 hours. The film thickness increased by a factor of 1.8 during the annealing, resulting in a final thickness of 400 nm, which

[c)] Electronic mail : eom@engr.wisc.edu

was confirmed by cross-sectional transmission electron microscopy (TEM). Atomic force microscopy (AFM) imaging revealed a smooth surface morphology with an RMS roughness of ~3 nm. The chemical composition of Film 1 was obtained using wavelength dispersive x-ray spectroscopy (WDS), showing a Mg:B:O:C atomic ratio of 34.1: 58.4: 4.3: 3.2, respectively. By assuming that the carbon resides on boron sites and oxygen consumes magnesium to form MgO, the Mg:B ratio of Film 1 was found to be 1:2.07, which is close to the $MgB_2$ stoichiometry.

The epitaxial relationships and the crystalline quality of the $MgB_2$ thin films were assessed by four-circle x-ray diffraction. Both Film 1 and Film 2 show very similar x-ray diffraction patterns except for a slightly broader rocking curve width in Film 1. Figure 1(a) shows a *q*-2*q* scan of an epitaxial $MgB_2$ thin film (Film 2) grown on a (0001) $Al_2O_3$ substrate. The only substantial $MgB_2$ peaks are the 0001 and 0002 reflections, which clearly shows that the $MgB_2$ is oriented with its *c*-axis normal to the substrate. The rocking curve full width at half maximum (FWHM) of the 0002 $MgB_2$ reflection is 0.54°, which indicates that the crystalline quality of the film is good. We also investigated the in-plane texture of the film by scanning an off-axis peak. Figure 1(b) shows the azimuthal *f*-scan of the $MgB_2$ $10\bar{1}1$ reflection. The significant intensities every 60° of this reflection confirm that the film contains a single hexagonal texture in the film plane. Furthermore, the $MgB_2$ reflections are rotated 30° in the basal plane with respect to the $Al_2O_3$ lattice, resulting in a relationship between the substrate and $MgB_2$ film of $[11\bar{2}0]MgB_2 \parallel [10\bar{1}0]Al_2O_3$. The measured FWHM of the azimuthal *f*–scan of the $10\bar{1}1$ reflection is 1.0°. The *c*-axis lattice parameter determined from normal *q*–2*q* scans is 3.52±0.01Å, which is the same as the bulk value[1].

The microstructure has been studied by cross-sectional TEM. Figure 2(a) is a low magnification bright field TEM image of a 4000 Å thick $MgB_2$ film grown on a (0001) $Al_2O_3$ substrate. Figure 2(b) and (c) are the selected-area electron diffraction (SAED) pattern taken from the film and the substrate, respectively. Epitaxial growth of $MgB_2$ is evident and no grain boundaries are seen in the film. The epitaxial orientation relationship determined by comparing the SAED pattern of the film and the substrate is in accordance with that determined by the x-ray studies. The high-resolution TEM (HRTEM) image in Fig. 2(d) shows distinct interface layers of $MgAl_2O_4$ and MgO between the $MgB_2$ and $Al_2O_3$. The HRTEM image clearly shows that both $MgAl_2O_4$ and MgO grow epitaxially on the (0001) $Al_2O_3$, with an orientation relationship of $(111)[1\bar{1}0]$ MgO//$(111)[1\bar{1}0]$ $MgAl_2O_4$//$(0001)[10\bar{1}0]$ $Al_2O_3$. The details of TEM studies will be given elsewhere[13].

The transition temperature was measured with a SQUID magnetometer in a magnetic field of 5 mT, applied parallel to the film surface. Figure 3 shows an extremely sharp transition with onset at 35 K with full shielding. The 10% to 90% width of the inductive transition is ~1 K, which is similar to inductive transitions for the best bulk and single crystal samples made so far[14].

The resistance was measured by a standard four-point technique in magnetic fields up to 9 T as a function of temperature. Figure 4 shows the zero field resistive transition, which indicates ρ(40 K) = 6.5 μΩcm and a residual resistance ratio ( RRR ) of ~2. Figure 5 shows the infield transitions for field applied parallel to the c-axis and to the ab plane of the film, and at this larger scale a partial transition at zero field is seen at 38 K. The in-field resistive transitions, with the possible exception of the 1 T parallel field transition, exhibit very little broadening up to the highest field measured (9 T), unlike our earlier measurements on fiber-textured films[3]. The upper critical field was defined for parallel ($H_{c2}^{\parallel}$) and perpendicular ($H_{c2}^{\perp}$) field by extrapolating the steep part of the transition to the normal state resistance[3]. The inset to Figure 4 shows the upper critical field vs. temperature. The anisotropy ratio, $H_{c2}^{\parallel} / H_{c2}^{\perp}$, is about 3 at 24 K, rather greater than film[3,15] and single crystal values[14]. At low temperatures, the nearly parallel trends and similar slopes of $H_{c2}^{\parallel}$(T) and $H_{c2}^{\perp}$(T) suggest decreasing anisotropy with decreasing temperature, a trend which is opposite to some recent data[16,17]. It is clear that there is as yet no convergence on what the upper critical field anisotropy of $MgB_2$ is.

The critical current density, $J_c$, was determined by magnetization measurements using a vibrating sample magnetometer in the field range of 0 to 12 T. Figure 6 shows $J_c$ vs. magnetic field at 4.2 and 20K in perpendicular magnetic field. To estimate $J_c$, we used the standard expression for the critical state of a thin film with rectangular area[18], $J_c = 2\Delta M(12b)/(3b-d)d$, where *b* = 4.5 mm and *d* = 3.0 mm are the film dimensions and $\Delta M$ is the total magnetization hysteresis measured from the up and the down branches of the magnetization curve. This analysis yields $J_c$ values of $4.5\times10^6$ A/cm$^2$ at 4.2 K and 1 T, and $1\times10^6$ A/cm$^2$ at 20 K and 1 T. These values contrast the almost reversible behavior of bulk single crystals[14] and are significantly larger than the values obtained from polycrystalline bulk forms of $MgB_2$ [2,19,20], where grain boundary flux pinning is often thought to be operating. In this case grain boundaries cannot be responsible for the flux pinning but it is evidently still very strong. We assume that some site interchange disorders must be contributing to the flux pinning.

In conclusion, we have demonstrated the growth of epitaxial $MgB_2$ films with high crystalline quality and high $J_c$. Although the films are epitaxial the properties are distinctively different from bulk single crystals in terms of their very strong flux pinning and suppressed $T_c$. We attribute this to alloying of the $MgB_2$ film by impurity atoms or some other defects. The growth of epitaxial $MgB_2$ thin films with controlled orientations and properties opens a new avenue to understand the superconductivity in $MgB_2$ and potential applications for both electronic devices and high field magnets.

The work at the University of Wisconsin was supported by National Science Foundation through the MRSEC for Nanostructure Materials. We are grateful to A. Gurevich for discussions. X.Q.P gratefully acknowledges the financial support of the National Science Foundation through grant DMR 9875405 (CAREER) and DMR/IMR 9704175.

Figure Captions:

Fig. 1. X-ray diffraction scans of an epitaxial $MgB_2$ thin film grown on a (0001) $Al_2O_3$ substrate (a) $\theta$-$2\theta$ scan and (b) $\phi$-scan of the $10\bar{1}1$ $MgB_2$ reflection. . The $Al_2O_3$ substrate peaks are marked as *. $\phi\equiv0°$ is aligned to be parallel to the $[11\bar{2}0]$ in-plane direction of the $Al_2O_3$ substrate. The FWHM of the 0002 $MgB_2$ peak is 0.28° in $2\theta$ and 0.54° in $\omega$ (rocking curve). The FWHM of the $10\bar{1}1$ $MgB_2$ reflection in $\phi$ is 1.0°. These scans indicate that the lattice constants of this $MgB_2$ film (Film 2) are $a$ = 3.08±0.02 Å and $c$ = 3.52±0.01 Å.

Fig. 2. (a) Bright-field cross-sectional TEM image of 4000 A thick $MgB_2$ thin film grown on a (0001) $Al_2O_3$ substrate, (b) SAED of $MgB_2$ along the $[11\bar{2}0]$ zone axis, (c) SAED of $Al_2O_3$ substrate along the $[10\bar{1}0]$ zone axis, and (d) cross-sectional HRTEM micrographs of epitaxial $MgB_2$ thin films on a (0001) $Al_2O_3$ near the substrate.

Fig. 3 $T_c$ data obtained with a SQUID magnetometer utilizing a 5 mT magnetic field, applied parallel to the film surface.

Fig. 4. Resistivity vs. temperature of $MgB_2$ thin film. The inset shows the $H_{c2}$ for fields applied parallel and perpendicular to the crystal axes.

Fig. 5. Resistive transitions in field for the film in fields up to 9 T. $H_{c2}(T)$ is defined by the extrapolation of the normal state resistance and the steep loss of resistance and is shown in the inset to Figure 4.

Fig. 6 Critical current density vs. magnetic field measured using a vibrating sample magnetometer at 4.2 and 20K in perpendicular magnetic field. Deviations below 1T are associated with the lack of full flux penetration and actual $J_c$ values are higher.

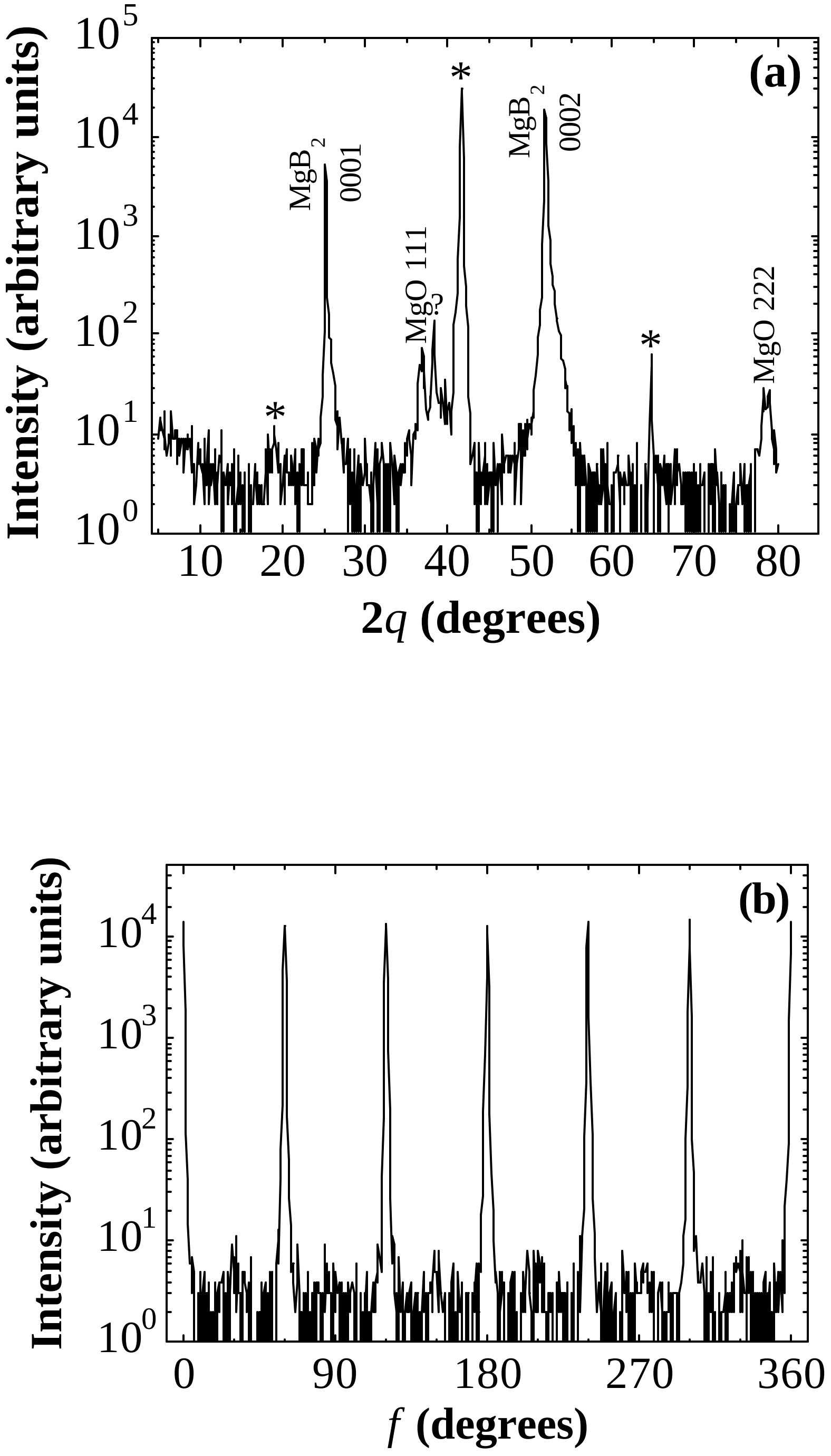

Figure 1

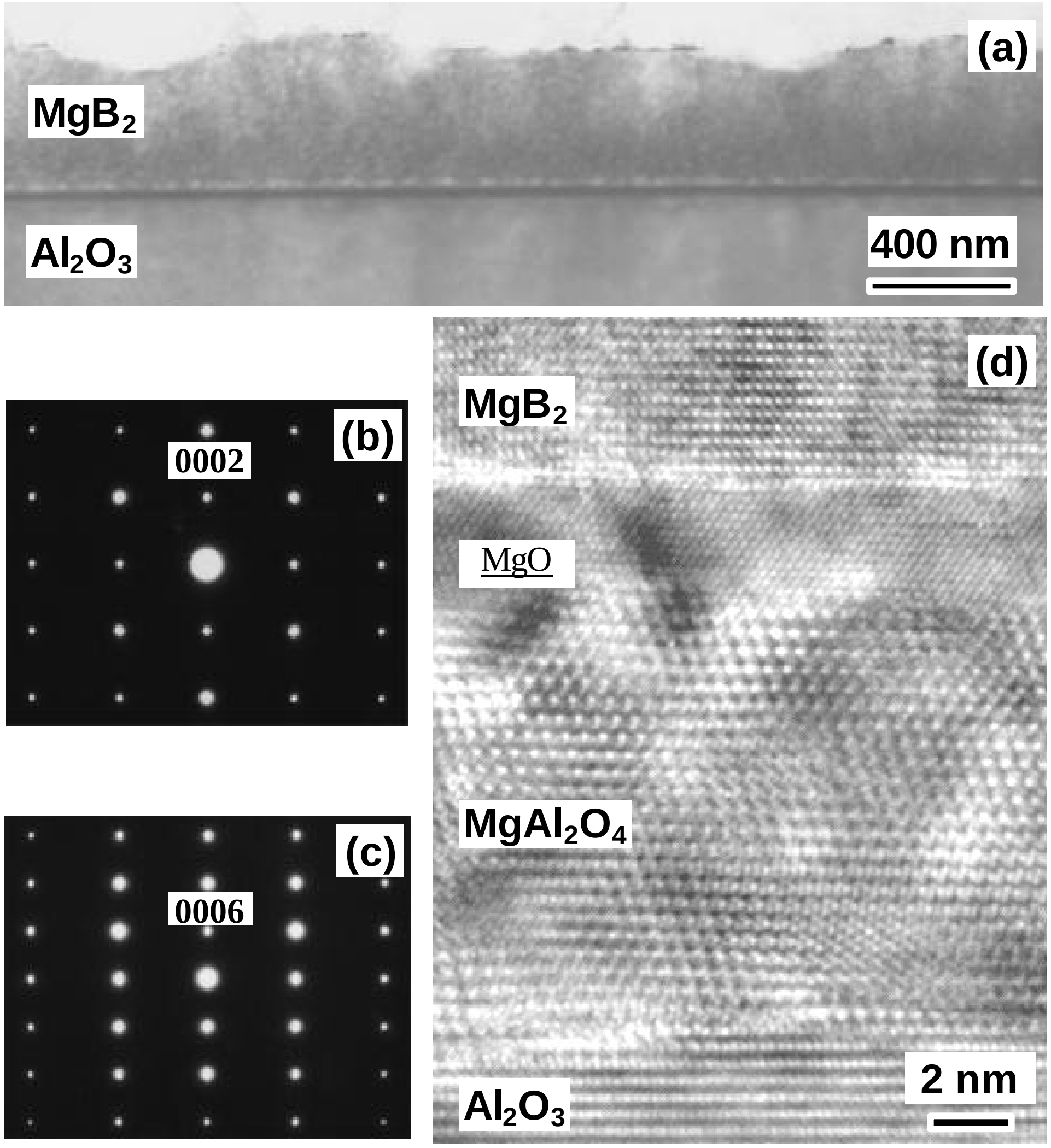

Figure 2

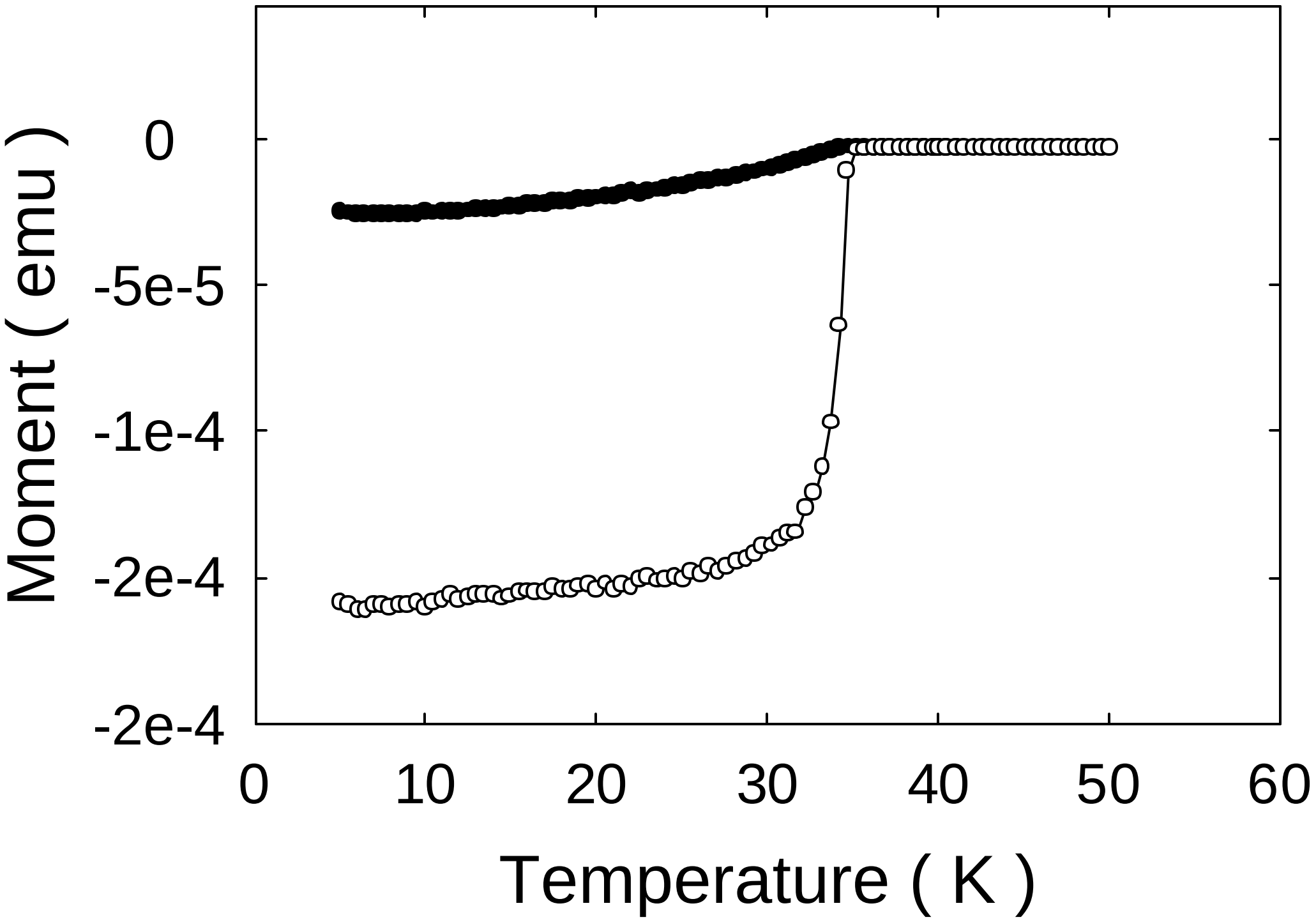

Figure 3

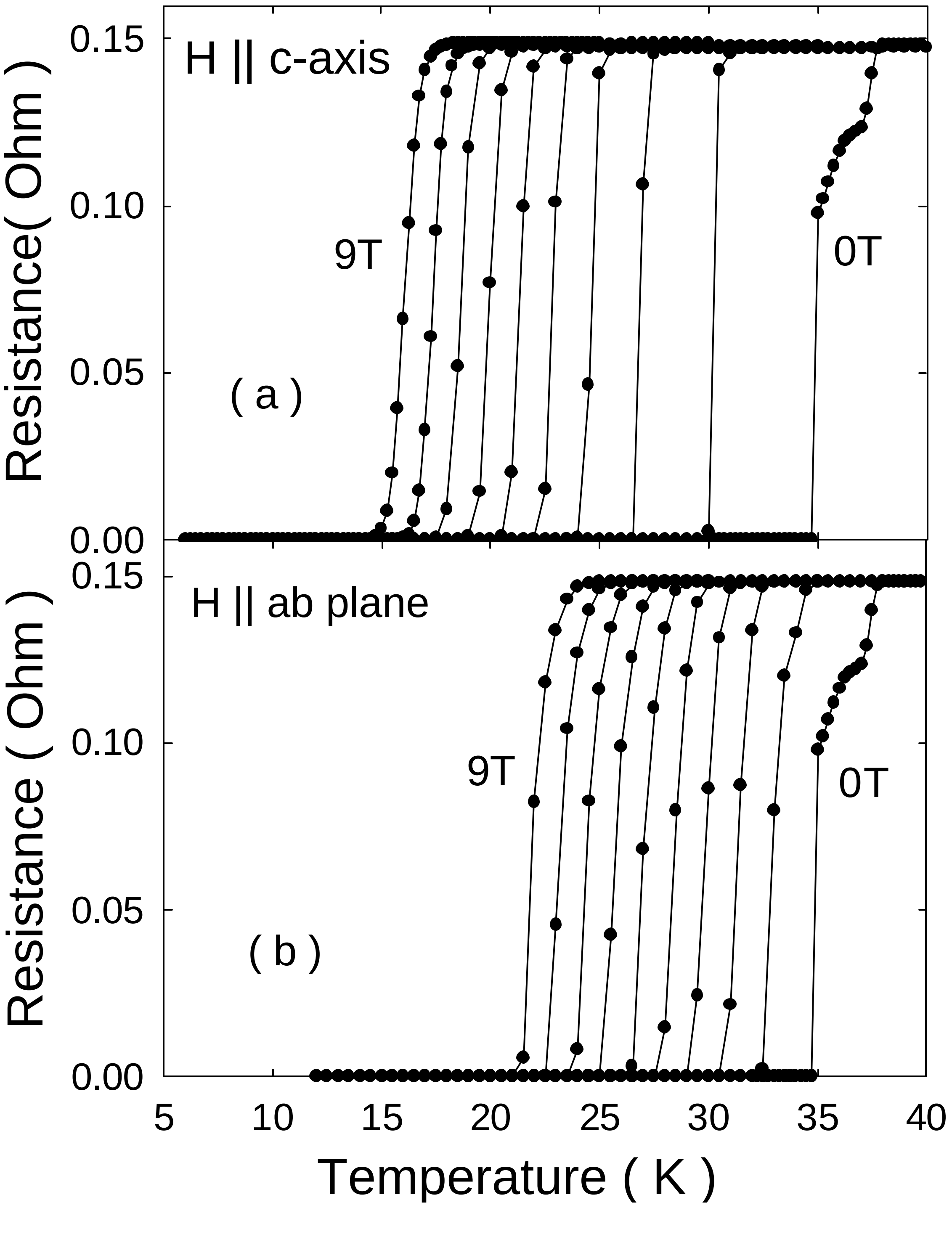

H || c-axis
9T
0T
( a )
Resistance( Ohm )
H || ab plane
9T
0T
( b )
Resistance ( Ohm )
Temperature ( K )

Figure 4

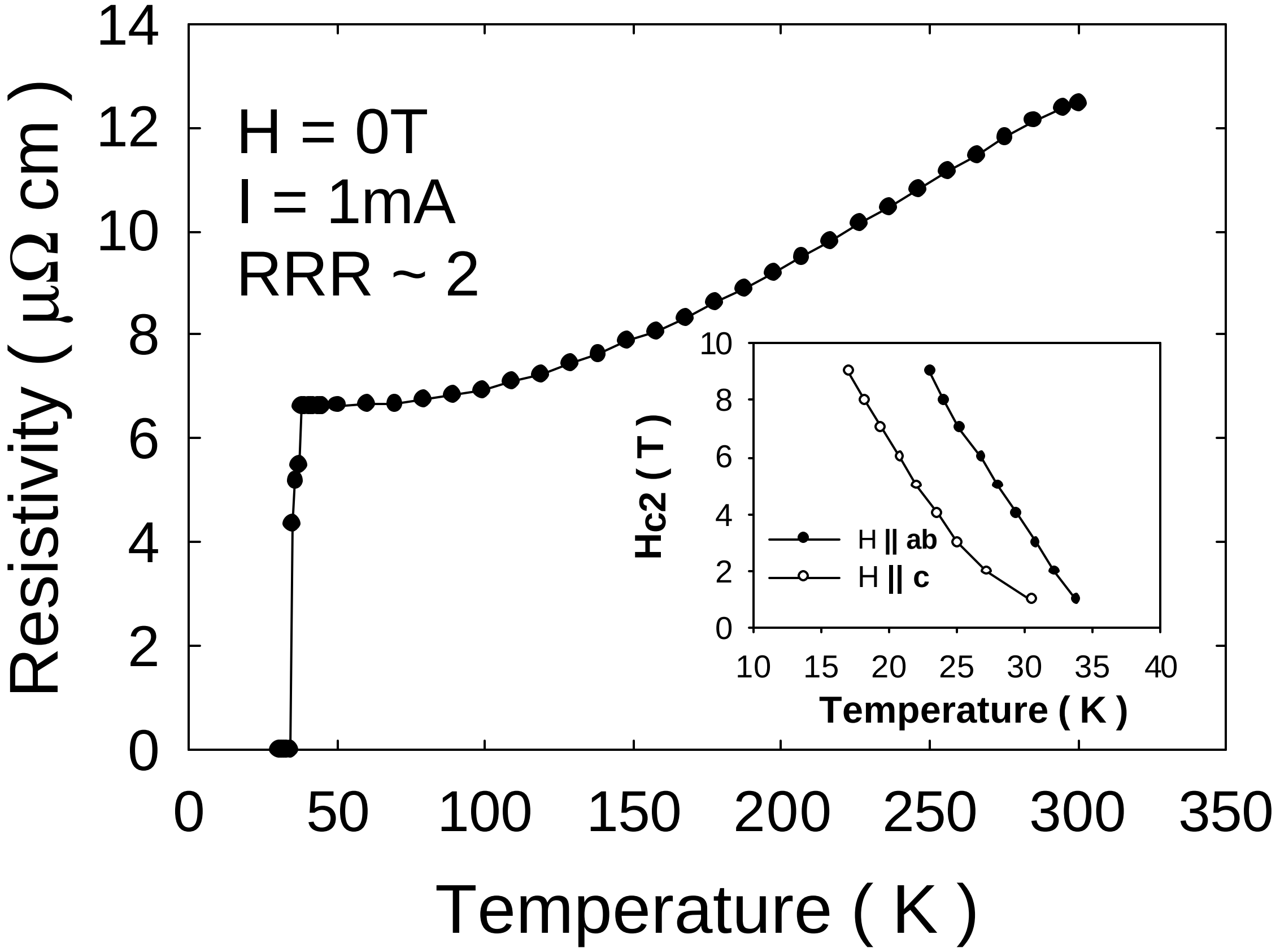

Figure 5

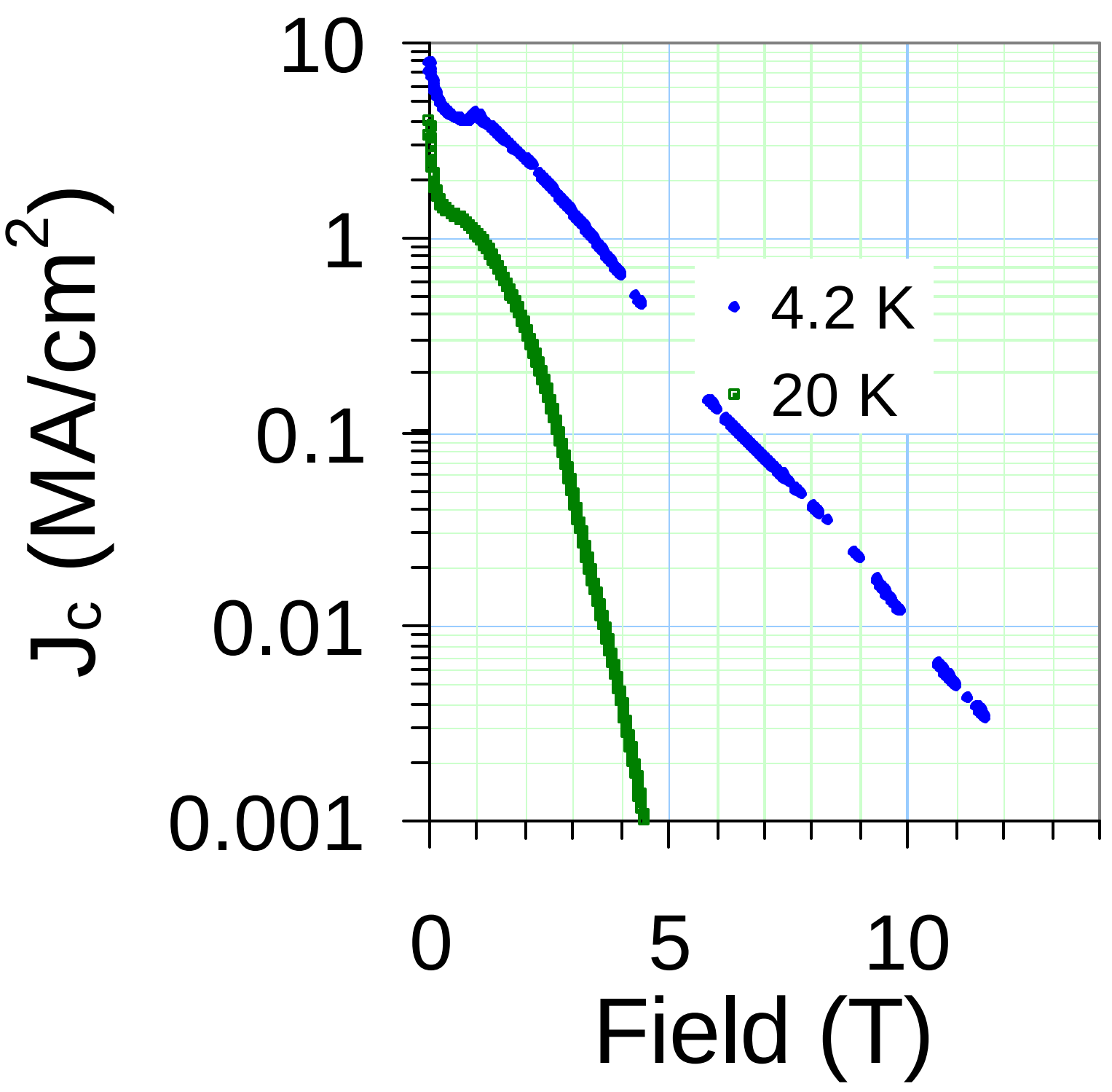

Figure 6